\documentclass[usenatbib]{mn2e}
\usepackage{graphicx}
\usepackage{amsmath}
\usepackage{amssymb}
\usepackage{times}
\usepackage[draft]{hyperref}
\pdfoutput=1

\title[Optical and NIR velocity dispersions of early-type galaxies]{Optical and near-infrared velocity dispersions of early-type galaxies\thanks{Based on new observations collected at the European Southern Observatory, Cerro Paranal, Chile; ESO program 082.B-0897A.}}  \author[J. Vanderbeke et al.]
{Joachim Vanderbeke$^{1,2}$, Maarten Baes$^1$, Aaron J. Romanowsky$^{3,4}$, Linda Schmidtobreick$^2$ \\
  $^{1}$ Sterrenkundig Observatorium, Universiteit Gent, Krijgslaan 281
  S9, B-9000 Gent, Belgium\\
$^{2}$ European Southern Observatory, Alonso de C\'{o}rdova 3107, Vitacura, Santiago, Chile\\
$^{3}$ UCO/Lick Observatory, University of California, Santa Cruz, CA 95064, USA \\
$^{4}$ Departamento de Astronom\'{\i}a, Universidad de Concepci\'on, Casilla 160-C, Concepci\'on, Chile}

\begin{document}

\date{Accepted 2010 November 16. Received 2010 November 15; in original form 2010 August 31}

\maketitle

\label{firstpage}

\begin{abstract}
  We have carried out a systematic, homogeneous comparison of optical and near-infrared dispersions. Our magnitude-limited sample of early-type galaxies in the Fornax cluster comprises 11 elliptical and 11 lenticular galaxies more luminous than $M_B = -17$. We were able to determine the central dispersions based on the near-infrared CO absorption band head for 19 of those galaxies. The velocity dispersions range from less than 70 km s$^{-1}$ to over 400 km s$^{-1}$. We compare our near-infrared velocity dispersions to the optical dispersions measured by \cite{Kuntschner2000}. Contrary to previous studies, we find a one-to-one correspondence with a median fractional difference of 6.4\%. We examine the correlation between the relative dust mass and the fractional difference of the velocity dispersions, but find no significant trend. Our results suggest that early-type galaxies are largely optically thin, which is consistent with recent Herschel observations.  \end{abstract}


\section{Introduction}

The dust content of early-type galaxies (ellipticals and lenticulars), and the effects on galaxy parameters inferred from optical observations, is still unclear and a hot topic of debate.  Dust in early-type galaxies was first observed in the form of dust lanes and patches \citep{Hawarden1981,Ebneter1985,Veroncetty1988}. Later HST observations revealed that dust extinction features exist in a large fraction of early-type galaxies \citep{vanDokkum1995}. Nonetheless, the optical dust features could not explain the high far-infrared (FIR) fluxes found by \cite{Goudfrooij1995} (based on \textit{Infrared Astronomical Satellite} (IRAS) observations). Their results implied dust masses exceeding the values from optical extinction by nearly an order of magnitude, indicating that the major part of the dust is diffusely distributed. Furthermore, the dust mass estimates based on IRAS flux densities are only a lower limit for the true dust masses, because IRAS is not sensitive to cold dust.  FIR observations of early-type galaxies selected from the \textit{Infrared Space Observatory} (ISO) archive found that the colder dust component dominates the total dust mass, which is typically more than 10 times larger than the dust masses previously estimated using IRAS observations \citep{Temi2004}. Recent FIR Spitzer observations of elliptical galaxies show evidence of diffuse dust \citep{Temi2007} and ground-based submillimetre continuum observations with SCUBA have already revealed that galaxies indeed contain large amounts of cold dust, but that the submm emission of some of the elliptical galaxies may be synchrotron rather than dust emission \citep{Vlahakis2005}. Nevertheless, the Herschel Space Observatory is crucial for detecting the missing cold dust component and, as a consequence, for making accurate dust estimates in ellipticals \citep{Boselli2010}. Science Demonstration Phase results for the Herschel Virgo Cluster Survey \citep{Davies2010} did not confirm the diffuse dust thesis in elliptical galaxies: \cite{Clemens2010} found no detection for passively evolving early-type galaxies and \cite{Baes2010} did not find evidence for a diffuse dust component in M87, explaining the FIR emission by a synchrotron model.

The influence of dust on the optical and NIR photometry of early-type galaxies was studied by \cite{Michard2005} and models to predict the effect of dust were developed by \cite{Witt1992} and \cite{Wise1996,Wise1997}. \cite{Baes2000,Baes2001,Baes2002} and \cite{Baes2000b} were the first to include the influence of dust on the observed stellar kinematics in their models. They showed that dust may bias optical observations through absorption and scattering, thus influencing the photometric and kinematic data.  Such effects would have widespread ramifications for studies of early-type galaxies, as velocity dispersions play a significant role in tracking the mass evolution of early-type galaxies \citep{Vandermarel2007}, and appear in empirical relationships such as the the $\text{M}_{\text{BH}}$-$\sigma$ relation \citep{Ferrarese2000,Gebhardt2000,Graham2008,Gultekin2009,Kormendy2009,Graham2010}, the Faber-Jackson relation and the Fundamental Plane relation \citep{Djorgovski1987,Dressler1987,Guzman1993,Pahre1995,Prugniel1996,Graham1997,Gavazzi1999,Bernardi2003,Michard2004,Desroches2007,FraixBurnet2010,Ribeiro2010}.

One way to investigate the effect of dust on the kinematics is to compare the optical and NIR velocity dispersions. The $K$-band is the optimal region to perform a study on NIR velocity dispersions: the extinction by dust in the $K$-band is only 7\% of that in the $B$-band (where dust is very opaque) and the wavelengths are short enough to evade dilution of the stellar continuum by hot dust \citep{Gaffney1995}. In the $K$-band, one can use the 2.29~$\mu$m (2$-$0) $^{12}$CO absorption band head as a kinematical tracer. It is the strongest absorption feature in galactic spectra in the 1$-$3 $\mu$m range and increases in strength with decreasing effective temperature or increasing radius of the underlying stars \citep{Silge2003}. This late-type star feature can be used to probe the kinematics of the red stellar population of galaxies \citep{Gaffney1993} and is intrinsically sharp and deep, spectrally isolated from other strong absorption and emission features and is located in a dark part of the infrared sky spectrum \citep{Lester1994}. The advent of efficient NIR spectrographs and detectors on large telescopes has made the CO absorption band a generally available tool for kinematic studies (e.g. \citealt{Silge2005,Nowak2007,Nowak2008,Lyubenova2008}).

Comparative studies about NIR and optical velocity dispersions were performed in the past. \cite{Silge2003} were the first to perform a systematic study, but their sample was biased to lenticulars (it contained 25 galaxies, of which 7 ellipticals and 18 lenticulars) and was based on inhomogeneous optical velocity dispersions, obtained from different papers with different instruments, methods and extraction windows. They found that the velocity dispersion decreases with wavelength, opposite to the theoretical expectations of \cite{Baes2002}. \cite{Silva2008} studied the stellar populations in early-type galaxies, using the strong spectral features near 2.2 $\mu$m. They found a one-to-one correspondence between the optical and NIR velocity dispersions, based on a homogeneous set of 4 lenticular and 7 elliptical galaxies. 10 out of the 11 galaxies are part of our new, extended sample, in which we excluded NGC\,1344, because this galaxy was not included in the sample of \cite{Kuntschner2000}. \cite{Rothberg2010} inspected the NIR and optical velocity dispersions in infrared-luminous mergers. Their reference sample of elliptical galaxies did not show much evidence for a $\sigma$-discrepancy, but they found a strong correlation with IR-luminosity and dust masses in merging galaxies.

To investigate the disagreement between the different velocity dispersion studies, we embarked on a project, for the first time using a complete and well-balanced sample of 22 early-type galaxies (comprising 11 ellipticals and 11 lenticulars). All galaxies are members of the Fornax cluster, the nearest galaxy cluster after the Virgo cluster. Fornax is considerably more compact and regular in shape than Virgo, doubling the central density of galaxies but having a total mass of nearly an order of magnitude lower. This makes the Fornax cluster a good representative of the groups and poor clusters in which most galaxies in the universe reside \citep{Jordan2007}. Nevertheless, the main reason for the choice of our sample was the availability of uniform spectroscopic optical data, obtained by \cite{Kuntschner2000}. In his paper, velocity dispersion, age, metallicity and line strength are discussed for all galaxies in the sample.

This paper is organized as follows: \S~\ref{sec:data} presents the data and the data reduction, \S~\ref{sec:results} presents the comparison of the optical and IR velocity dispersions and the relation with the dust mass. In \S~\ref{sec:conclusions} our conclusions are presented.  

\section{Observations and data reduction}
\label{sec:data} 

\subsection{Data sample} 

This study is based on the complete magnitude-limited sample of \cite{Kuntschner2000}. This sample, presented in Table \ref{galaxies}, has been selected from the catalogue of Fornax galaxies of \cite{Ferguson1989}, in order to obtain a complete sample down to $B_T=14.2$ or $M_B = -17$ and contains 11 elliptical and 11 lenticular galaxies. Recent measurements of $B_T$ were taken from the NASA/IPAC Extragalactic Database.

\subsection{Observations}

The observations were performed between 20 October 2008 and 26 January 2009 with the VLT using the SW arm of ISAAC \citep{Moorwood1998} in spectroscopic medium resolution mode (SWS-MR). The characteristics of the detector and the instrumental set-up are given in Table \ref{tabinstrsetup}.

We used the nod-on-slit mode (double subtraction technique). Every observational sequence was started with the galaxy centered on the slit near one end and an individual spectrum was taken. The galaxy was then moved $60''$ or $90''$ towards the other end of the slit and two more integrations were executed. Then the galaxy was placed again at the original slit position where another spectrum was taken. This ABBA sequence was repeated a number of times, resulting in multiple individual two-dimensional spectroscopic images. In order to remove properly the sky background for the observations of big and bright galaxies (i.e. for NGC\,1399, NGC\,1316, NGC\,1380, NGC\,1404 and NGC\,1427 in particular), we observed separate sky spectra. That way, only half of the frames contained the galaxy spectrum, resulting in an observing sequence of the form OSSO (O=object integration, S=sky integration). Total on source exposure times were between 900 and 2400s [see Table \ref{galaxies} for a detailed listing, with `(1/2)' denoting OSSO observations]. The seeing was generally sub-arcsecond, reaching as low as $\sim 0.5''$ in the best cases.

To be able to remove the telluric lines, we observed B-type standard stars with the same observing sequence we have used for our science targets. These hot stars do not have any spectral features in the wavelength interval we are considering. Moreover, their continuum in the $K$-band is well approximated by the Rayleigh-Jeans part of the blackbody spectrum \citep{Silva2008}. To remove the stellar signature, we divide the spectrum by a template of the corresponding type of star \citep{Pickles1998}. All the remaining variation in the spectra is due to observational features and is caused by the telluric lines and the instrument response, so, by dividing the galaxy spectra by this residual template spectrum, we remove the telluric lines and the detector signature.  Because of the variability of the NIR sky, we impose the conditions that the difference in observing time between the target and the standard star has to be less then 2 hours and that the difference in airmass between the observations has to be less than 0.2.

\begin{table*}
\caption{\label{galaxies} Galaxy sample. }
\begin{tabular}{lccccccccc}
\hline
Galaxy & Type  & $B_T$ & Exposure time & PA &  $\sigma_{\text{opt}}$ & $\sigma_{\text{NIR}}$ & $\sigma_{\text{NIR,23 templates}}$ &EW$_{\text{CO}}$ & S/N \\
&& [mag] & [s] & [$^{\circ}$]&[km s$^{-1}$]& [km s$^{-1}$] & [km s$^{-1}$] &[\AA]&\\
\hline
NGC\,1316 & S0 pec &9.4 &1800 (1/2)&47 & 221.0 $\pm$ 11.0 & 237.7 $\pm$ 11.1&$218.6\pm9.9$&13.56&89 \\
NGC\,1336 & E4&13.1&2400&10 & 96.0 $\pm $ 5.0& 119.0 $\pm$ 8.2&$126.9\pm  7.0$&13.04&66\\
NGC\,1339&E5&12.5&1200&159&158.0  $\pm$ 8.0 &182.4 $\pm$ 9.2& $175.2\pm 9.4$&14.41&57\\
NGC\,1351&E5&12.5&1200&140&157.0  $\pm$ 8.0&153.0  $\pm$ 6.6 & $148.4\pm 5.5$&9.58&73\\
NGC\,1373&E3&14.1&2400&130&75.0  $\pm$ 4.0 &79.8  $\pm$ 4.7& $ 72.5\pm 4.0  $&11.26&66\\
NGC\,1374&E0&12.0&1200&120&185.0  $\pm$ 9.0&206.8  $\pm$ 10.0& $       193.9\pm9.7 $&13.72&72\\
NGC\,1375&S0&13.2&1800&90&56.0  $\pm$ 10.0 &64.1  $\pm$ 4.2&$       58.2\pm  3.8 $&11.66&68\\
NGC\,1379&E0&11.8&1200&65&130.0  $\pm$ 7.0 &130.0  $\pm$ 6.8&$       123.3\pm  5.3  $&12.02&90\\
NGC\,1380&S0&10.9&1800 (1/2)&7&219.0 $\pm$ 11.0&189.8 $\pm$ 16.6& $       199.8\pm  20.8 $&12.19&59\\
NGC\,1380A&S0&13.3&2400&178&55.0 $\pm$ 9.0&59.9 $\pm$ 9.1&$       51.2\pm   8.6    $&11.41&33\\
NGC\,1381&S0&12.4&1200&139&153.0 $\pm$ 8.0&155.2 $\pm$ 5.7&$       148.3\pm  4.6    $&11.70&129\\
NGC\,1399&E0, cD&10.6&1800 (1/2)&175&375.0 $\pm$19.0&405.5 $\pm$ 33.1&$       409.0\pm  34.2     $&11.87&40\\
NGC\,1404&E2&11.0&1800 (1/2)&163&260.0 $\pm$ 13.0&246.9 $\pm$ 21.6& $       240.9\pm   19.0    $&13.30&84\\
NGC\,1419&E0&13.5&2400&125&117.0 $\pm$ 6.0&125.4 $\pm$ 5.0&$       119.2\pm   3.7    $&10.45&108\\
NGC\,1427&E4&11.8&1800 (1/2)&76&175.0 $\pm$ 9.0&154.9 $\pm$ 17.5&$       150.3\pm   14.6   $&9.87&62\\
IC\,1963 &S0&12.9&2400&83 &58.0 $\pm$10.0&48.6 $\pm$ 5.6&$       40.3\pm  4.7   $&9.97&57\\
IC\,2006&E&12.2&1800&10&136.0 $\pm$ 7.0&125.4 $\pm$10.2&$       118.7\pm  8.1     $&14.18&50\\
ESO\,359-G02&S0&14.2&1800&50&45.0 $\pm$ 8.0&...&...&...&7\\
ESO\,358-G06&S0&13.9&1800&31&57.9 $\pm$ 10.8&55.1 $\pm$ 25.0&$52.8\pm 24.8$&13.20&13\\
ESO\,358-G25&S0 pec&13.8&1800&60&58.0 $\pm$ 10.0&...&...&...&8\\
ESO\,358-G50&S0&13.9&1800&172&49.0 $\pm$ 8.0&...&...&...&17\\
ESO\,358-G59&S0&14.0&1800&160 &54.0 $\pm$ 9.0&70.0 $\pm$ 20.4&$       65.0\pm   14.9   $&10.83&29\\
\hline 
\end{tabular} \\
Notes.-Morphological type and optical velocity dispersions taken from \cite{Kuntschner2000}, $B_T$ from NASA/IPAC Extragalactic Database. 
Empirical S/N estimates were determined for the extracted spectra as described in \S \ref{sec:data_reduction}. 
\end{table*}

\begin{table}
\caption{\label{tabinstrsetup} The instrumental set-up.}
\begin{tabular}{ll}
\hline
Telescope     & VLT UT1/Antu          \\
Dates     & 20 October 2008 -  26 January 2009         \\
Instrument & ISAAC \\
\hline
Spectral range& $0.98 - 2.5\ \mu m$\\
Grating &    MR   \\
Dispersion  &   1.22 \AA\ pixel$^{-1}$   \\
Resolution (FWHM) &  8.196 \AA\ \\
Spatial Scale & 0.1484 arcsec pixel$^{-1}$ \\ 
Slit Width & 1 arcsec\\
Detector & Hawaii \\
Gain & 4.6 e$^-$ ADU$^{-1}$ \\
Read-out noise &$\sim 10$ e$^-$\\
Seeing & $< 1'' $\\
Pixel Size & 18.5$\mu m$\\
\hline
\end{tabular}
\end{table}

\subsection{Data reduction} 
\label{sec:data_reduction} 

\begin{figure}
\centering
\includegraphics[width=0.45\textwidth]{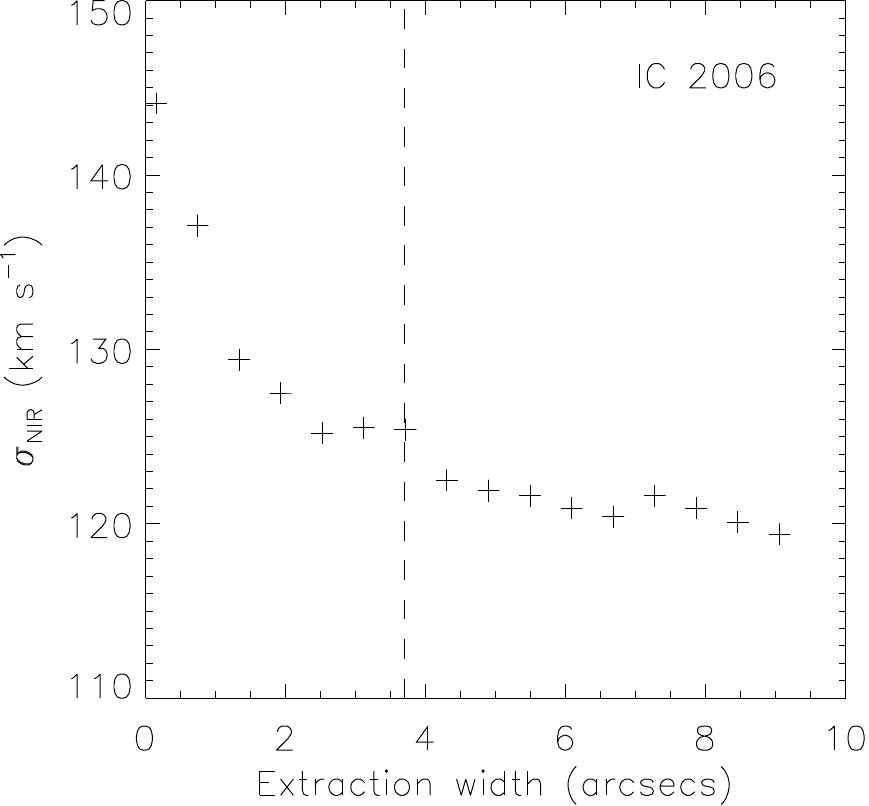}
\caption{Central velocity dispersion of IC\,2006 as a function of the extraction width. The dashed line shows the standard extraction aperture of 3.7 arcsecs.}
\label{dispersionextractionwindow}
\end{figure}

\begin{figure} \centering \includegraphics[width=0.45\textwidth]{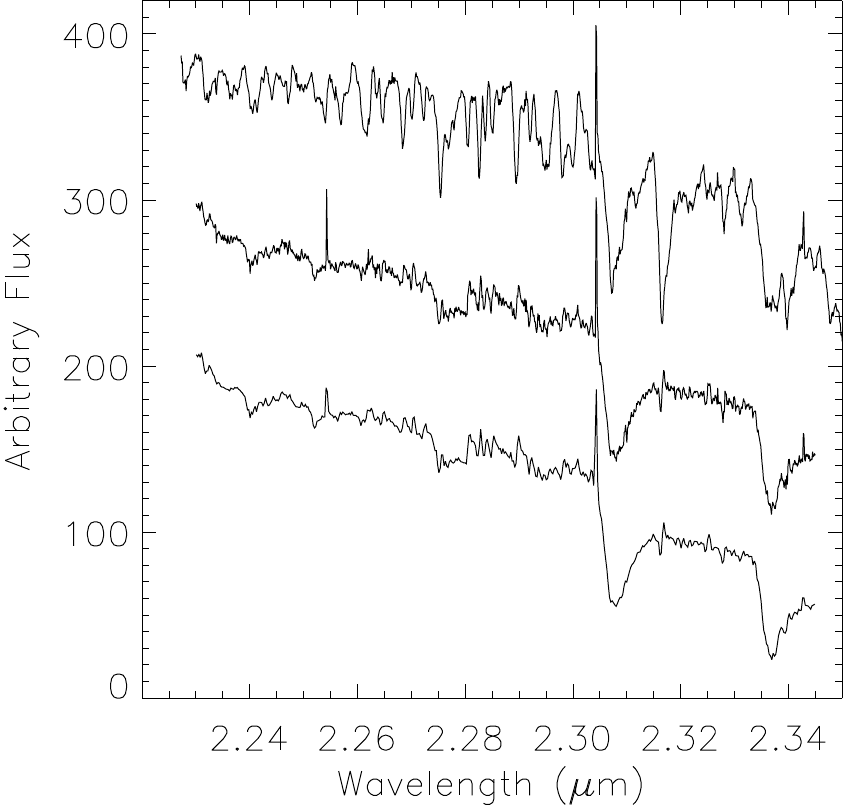} \caption{Data reduction steps applied to a central $3.7''$ extraction of NGC\,1381 are shown (from top to bottom) before corrections for telluric absorption (the pipeline result), after correction for telluric absorption and instrument sensitivity and after rebinning to $\sim$2.43\AA/pix. The sharp emission feature around 2.3 $\mu$m is a remnant of the telluric correction.No correction for redshift has been applied. The middle and the bottom spectra have been separated by an arbitrary additive offset; otherwise, they would be on top of each other.  }
\label{fig:data_reduction}
\end{figure}

The basic data reduction steps were performed with the ISAAC pipeline and MIDAS. The sky subtraction, bias subtraction, flat fielding and wavelength calibration (based on the OH-lines) were done by the ISAAC pipeline. The removal of telluric lines (using our standard star observations) was done with MIDAS. If possible and appropriate, we increased the S/N of the telluric profile by averaging different observations of telluric stars. For NGC\,1380, telluric lines have not been removed, because of a problem with the wavelength calibration of the standard star.  For all spectra we had some problem around 2.3~$\mu$m to correct a telluric absorption line with a very sharpe edge. As a result, a spike-like emission feature remains in the corrected spectra. We choose not to remove cosmic rays, but will ignore any contaminated pixel when fitting the stellar template.  An average of the 25 central rows was taken to obtain an effective aperture of $1 \times 3.7$ arcsec$^2$, choosing a slit width of 1 arcsec to obtain the needed spectral resolution and approximating the spatial width of the $2.3 \times 3.85$~arcsec$^2$ optical aperture used by \cite{Kuntschner2000}.  We do not increase the extraction width to obtain a better S/N, because the velocity dispersion can be strongly aperture-dependent (as shown in Fig.~\ref{dispersionextractionwindow}), and it is important to be well-matched to \cite{Kuntschner2000}. As a final step, the spectra were rebinned to a common wavelength increment (2.43 \AA/pix), approximately doubling the original step size. To illustrate the effect of the different data reduction steps, we refer to Fig.~\ref{fig:data_reduction}, where a central $3.7''$ extraction of NCG 1381 is shown for the different data reduction steps: the ISAAC pipeline result, after removing the telluric lines and after rebinning to the 2.43\AA/pix wavelength increment.  For each extracted spectrum, we have derived an empirical S/N following the method described by \cite{Stoehr2007}. The resulting S/N per rebinned element ranges from $\sim7$ for the faintest galaxies to more than 100 for the brightest ones and is listed in Table \ref{galaxies}.

The instrumental set-up, as described in Table \ref{tabinstrsetup}, gives a FWHM resolution of approximately 107 km s$^{-1}$, producing an instrumental contribution to the dispersion of $\sim45$ km s$^{-1}$.

\subsection{Velocity dispersion determination}
The sharp blue edge of the CO band head allows us to measure the kinematics accurately \citep{Silge2003}. There are several techniques to obtain the internal kinematical information, e.g. the Fourier correlation quotient (FCQ) method, developed by \cite{Bender1994} and used by \cite{Kuntschner2000}. We use the pPXF technique developed by \cite{Cappellari2004}. This method rebins the spectrum logarithmically and fits it directly in pixel space. We correct the template continuum shape using additive second degree Legendre polynomials and use a pure Gaussian to model the LOSVD. An alternative LOSVD is inspected in \S~\ref{sec:compNIRopt}, where Gauss-Hermite coefficients $h_3$ and $h_4$ are included to model the LOSVD. We apply pPXF to a wavelength range of $\sim2.23-2.345\, \mu$m, fitting not only the CO band head, but also including a part of the spectrum bluewards of it. To illustrate the effect of the velocity dispersion on the CO band head, we refer to Fig. 2 of \cite{Silge2003}. 

\begin{figure}
\centering
\includegraphics[width=0.45\textwidth]{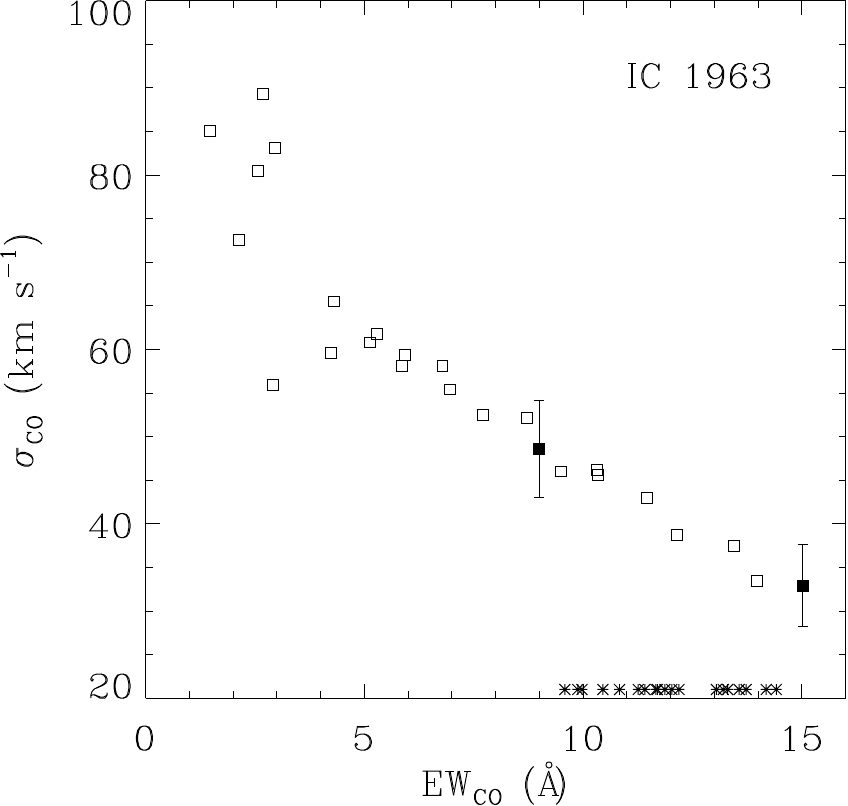}
\caption{Dispersion measured by the pPXF method for IC\,1963 as a function of the equivalent width of the input template star. The two filled squares are the results for an average template of K giants (EW$_{\text{CO}}=8.99$, $\sigma_0 = 48.6\pm 5.6$) and M giants (EW$_{\text{CO}}=15.01$, $\sigma_0 = 32.9 \pm 4.7$). The asterisks represent the EWs of the galaxies in the sample.}
\label{EW_sigma_IC1963}
\end{figure}

The choice of the template stars, which we obtained from the GNIRS and NIFS libraries \citep{Winge2009}, is an important aspect when determining the velocity dispersion: \cite{Silge2003} showed that the equivalent width of the template used for the fit is important, not the details of the spectral type. To account for the spectral resolution differences between ISAAC and the template libraries, we have logarithmically rebinned the latter to the same velocity scale as the ISAAC spectra, under the assumption that the shape of the instrumental spectral profiles can be well approximated by a Gaussian. Fig. \ref{EW_sigma_IC1963} presents the measured central velocity dispersion as a function of the equivalent width of the input template, with equivalent widths ranging from $\sim 1.5 \text{ to} \sim 15$. One concludes that the larger the EW of the CO band of the template, the lower the resulting velocity dispersion required to reproduce the galaxy spectrum. The relative values of the velocity dispersion are only comparable, if a fixed template is used to generate the fits. The filled squares are the results for an average template of K giants (HD206067, HD218594, HD39425 and HD4730 from the GNIRS library) and M giants (HD27796, HD30354, HD23574,  HD234791 from the NIFS library). To be consistent with \cite{Kuntschner2000}, we choose the average K giant spectrum to fit the galaxy spectra in a homogeneous way. An alternative template scheme is explored in \S ~\ref{sec:compNIRopt}, where velocity dispersions have been determined using all the templates of Fig.~\ref{EW_sigma_IC1963} as input templates for pPXF. The asterisks in Fig. \ref{EW_sigma_IC1963} represent the EWs of the galaxies in the sample, given in Table \ref{galaxies}. This shows that the stellar library spans the range of the observed CO strength in the galaxies.

Table ~\ref{galaxies} presents the resulting NIR velocity dispersion $\sigma_{\text{NIR}}$ and the optical velocity dispersions $\sigma_{\text{opt}}$ taken from \cite{Kuntschner2000}. The optical velocity dispersions have been used as the initial value in the pPXF method. After trying several approaches, we have used a bootstrap method to estimate the uncertainties on the velocity dispersions, resampling the residuals of the initial fit. For ESO\,359-G02, ESO\,358-G25 and ESO\,358-G50 we did not get a reliable fit, as a consequence of the low S/N for these galaxies and the low expected velocity dispersions (based on the optical values). We decided to exclude these galaxies from the sample. Afterall, their optical velocity dispersions are not reliable either; they suffered an instrumental broadening of $\sim 105$ km s$^{-1}$, which introduces optical systematic errors for $\sigma < 90$ km s$^{-1}$ [as stated in \cite{Kuntschner2000}]. 

\begin{figure*}
\centering
\includegraphics[]{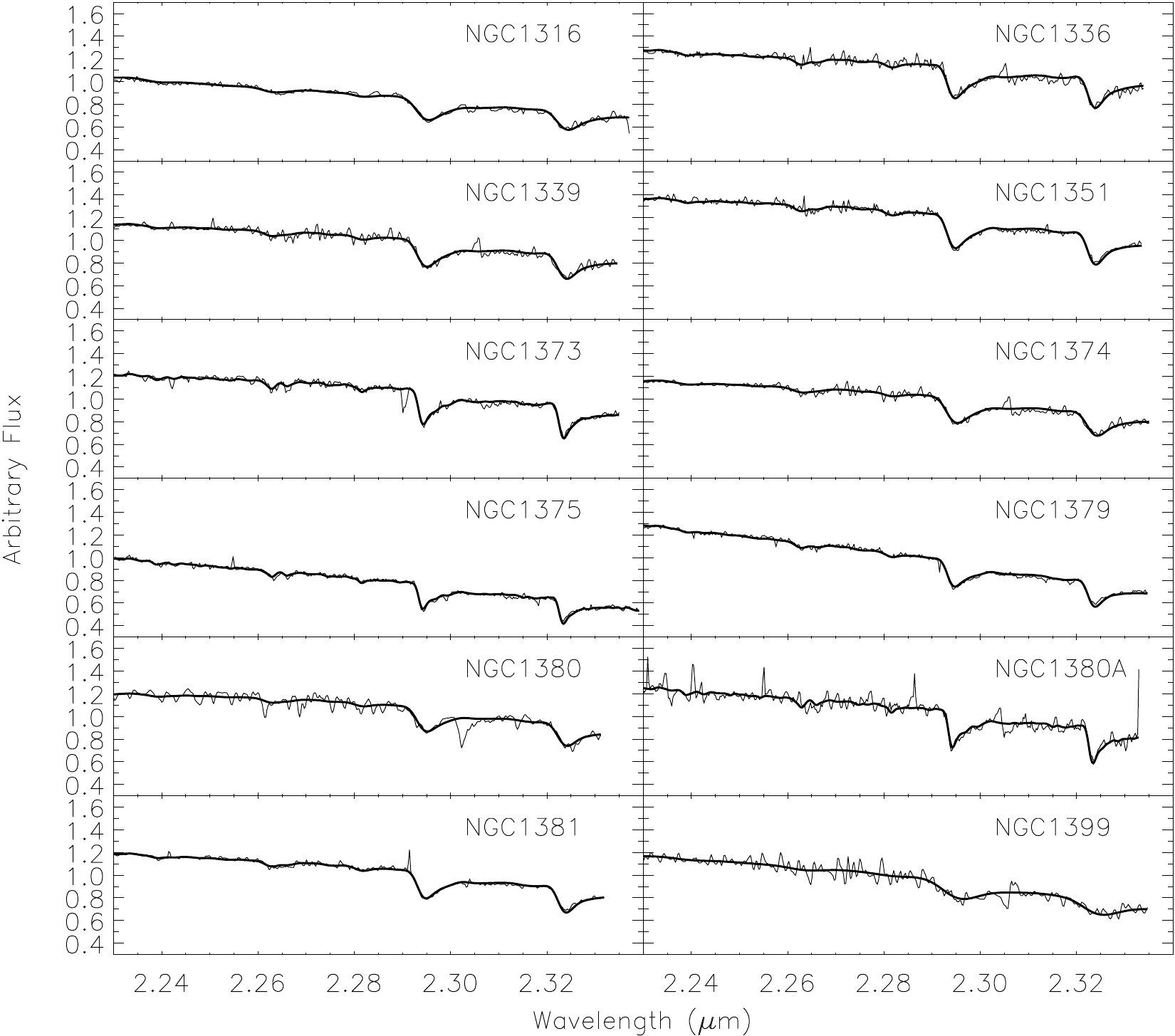}
\caption{Rest-frame spectra of galaxies (noisy thin curves) and spectra of the average KIII stellar spectrum convolved with the derived velocity distribution (smooth thick curves). Cosmic rays and poorly removed telluric features were flagged as bad pixels and not used to generate the fit. No removal of telluric lines was done for NGC\,1380.  } \label{fig_fit_galaxy_spectra}
\end{figure*}

\begin{figure*}
\centering
\includegraphics[]{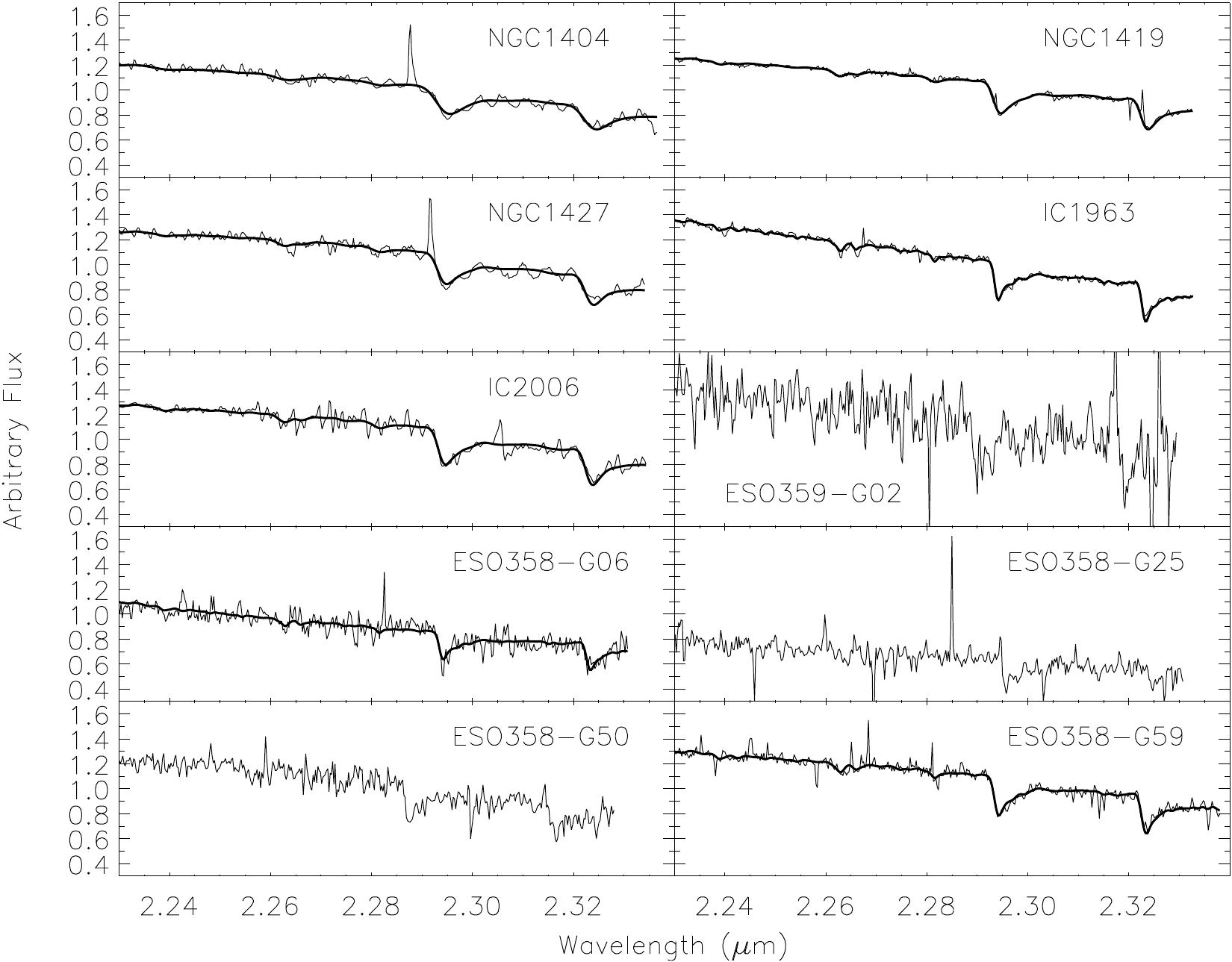}
\caption{Same as in Fig. \ref{fig_fit_galaxy_spectra}.}
\label{fig_fit_galaxy_spectra2}
\end{figure*}

The spectral fits are presented in Figs. \ref{fig_fit_galaxy_spectra} and \ref{fig_fit_galaxy_spectra2}: the thick smooth line is the resulting fit, the thin noisy line represents the galaxy spectrum. Cosmic rays, sharp features and poorly removed telluric features were flagged as bad pixels and were not used by the pPXF method to generate the fit. If we estimate the error on the fit by the dispersion of the residuals, we obtain $\chi^2/DOF<1$ for all galaxies, which indicates that we obtain good fits.

\section{Results and discussion} 
\label{sec:results}

\subsection{Comparing NIR velocity dispersions}

The line strength study of \cite{Silva2008} had 4 S0 and 6 E galaxies in common with our galaxy sample [of which 8 are observed with ISAAC and 2 (NGC\,1316 and NGC\,1399) with SINFONI at the VLT]. For their ISAAC observations in SWS-MR mode, they also used a $120''\times1''$ slit and had total exposure times of 3000s or 3200s. This resulted in extracted spectra with a S/N ranging from 48 to 280, based on the S/N determination of \cite{Stoehr2007}. However, they applied a procedure derived from the processing of optical long-slit spectra and did not use a simple AB subtraction, because this could cause inaccurate dark and background correction in the lower surface brightness parts of the luminosity profiles. They periodically observed velocity template standard stars, covering the range K5 III to M1 III. Those templates were fitted to the galaxy spectra applying the pPXF method, extracting the NIR velocity dispersions $\sigma_{\text{NIR,Silva}}$. Unfortunately, no uncertainties on the velocity dispersions were given. They used $1'' \times 1/8 R_e$ for the extraction window, which results in spatial widths ranging from 0.15 to 0.61 arcsec  (extracted $R_e$ for NGC\,1404 equals 2.9 arcsecs, kindly provided by Dr. Kuntschner, private communication). For the SINFONI observations, \cite{Silva2008} extracted a slit of $2''\times3''$ from the data cube. 

\begin{figure} \centering \includegraphics[width=0.45\textwidth]{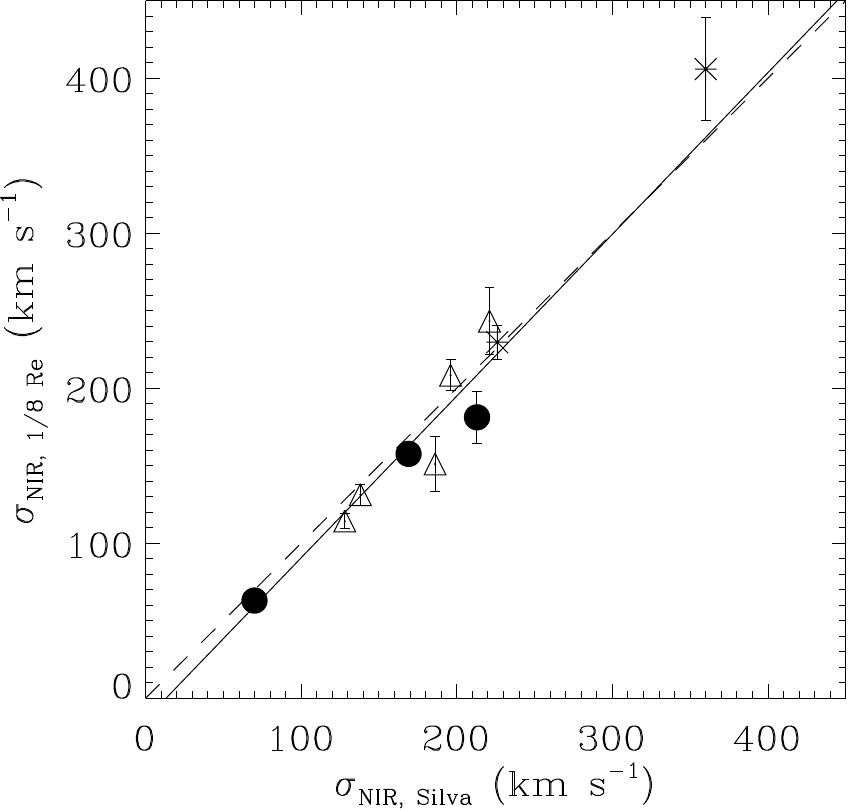} \caption{Correlation between the velocity dispersions determined with an extraction window of 1/8 $R_e$ and the NIR velocity dispersions from the literature \citep{Silva2008}. The dashed line has a slope of unity, the solid line is the best fit, given by equation \ref{eq:correxactSilvawithSinfoni1x3}. Here and in the following plots, the filled circles are S0 galaxies, while the open triangles are E galaxies. The two galaxies observed with SINFONI \citep{Silva2008} are represented with an asterisk. }
\label{corrCOexactsilva}
\end{figure}

In Fig. \ref{dispersionextractionwindow} it is shown that the velocity dispersion can depend on the extraction width. In order to compare our measurements with \cite{Silva2008}, we have redetermined the velocity dispersion for the 10 galaxies in common, adopting an extraction width of $1'' \times 1/8 R_e$ ($1''\times3''$ for NGC\,1316 and NGC\,1399) and fitting the average K giant template to the galaxy spectra. The measurements are denoted as $\sigma_{\text{NIR,1/8 Re}} $. The results are presented in Fig. \ref{corrCOexactsilva}, where the derived velocity dispersions $\sigma_{\text{NIR,1/8 Re}} $ are compared to the values $\sigma_{\text{NIR,Silva}}$, derived by \cite{Silva2008}.  In this and following figures, true elliptical galaxies are represented as open triangles, lenticular galaxies as filled circles. However, to distinguish the 2 galaxies observed with SINFONI by \cite{Silva2008}, we have marked those galaxies with
asterisks in Fig. \ref{corrCOexactsilva}. We have used the uncertainties on $\sigma_{\text{NIR}}$ as an estimate for the uncertainties
on $\sigma_{\text{NIR,1/8 Re}}$. The best-fitting line, which is given by
\begin{equation}
  \sigma_{\text{NIR,1/8 Re}}
  = 
  (1.04 \pm 0.05) \sigma_{\text{NIR,Silva}}
  - 
  (13.78 \pm 9.52), 
\label{eq:correxactSilvawithSinfoni1x3}
\end{equation}
has a slope consistent with 1, an intercept marginally inconsistent with 0 and a reduced $\chi^2$ of $1.89$. Using the velocity dispersions obtained with the fixed extraction width of $3.7''$ ($\sigma_{\text{NIR}}$) instead of $\sigma_{\text{NIR,1/8 Re}}$ does not change the fit significantly. Thus, our NIR results generally compare well to those of \cite{Silva2008} for the same galaxies.

\subsection{Comparing NIR and optical dispersions} \label{sec:compNIRopt}

\begin{figure}
\centering
\includegraphics[width=0.45\textwidth ]{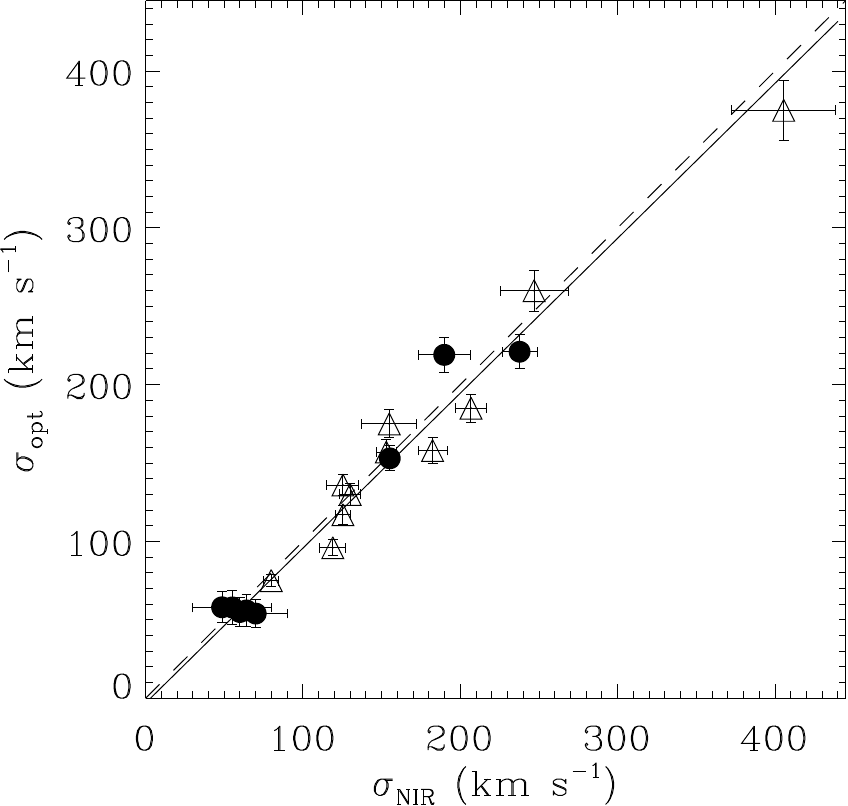}
\caption{Correlation between the dispersion measured from the CO band head and the optical dispersion \citep{Kuntschner2000}. The dashed line shows where two measurements are equal, the solid line is the best-fitting line, given by equation \ref{eq:corrCOopt}.  }
\label{corrCOoptical}
\end{figure}

Fig.~\ref{corrCOoptical} presents the correlation between the optical and NIR central velocity dispersions for 19 galaxies of our sample. The solid line 
\begin{equation}
  \sigma_{\text{opt}}
  = 
  (0.99 \pm 0.06) \sigma_{\text{NIR}} 
  - 
  (3.76 \pm 8.72)  
\label{eq:corrCOopt}
\end{equation}
shows the best fit, with a slope consistent with 1, an intercept consistent with 0 and a reduced $\chi^2$ of 1.09. The absence of a $\sigma$-discrepancy confirms the findings by \cite{Silva2008} and further generalizes their results due to the statistical completeness of this study, but does not agree with \cite{Silge2003}. Their best-fitting line had a slope of $1.189\pm0.084$ and an intercept of $-8.6\pm12.4$.

\cite{Kuntschner2000} had difficulties in determining the velocity dispersions for galaxies with $\sigma_{\rm opt}<70~\text{km s}^{-1}$ and noted that the results of $\sim50-60~\text{km s}^{-1}$ are only rough estimates. However, excluding the galaxies with $\sigma_{\rm opt}<70~\text{km s}^{-1}$ does not change the results significantly. 

\begin{figure}
\centering
\includegraphics[width=0.425\textwidth]{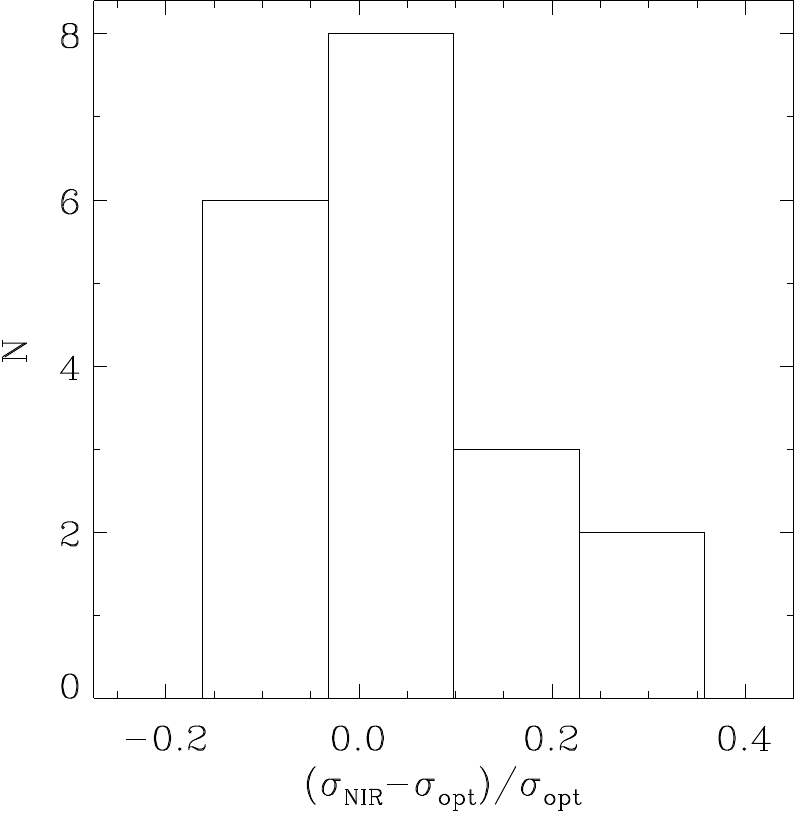}
\caption{Histogram showing the number of galaxies in each bin of fractional difference between infrared and optical measurements. The median fractional difference is 6.4\%, the mean fractional difference is 3.9\%.}
\label{histogramfracdiff}
\end{figure}

Fig.~\ref{histogramfracdiff} presents a histogram of the fractional difference between NIR and optical measurements of dispersion. The median fractional difference between the optical and the NIR velocity dispersions is 6.4\%, the mean fractional difference is 3.9\%. \cite{Silge2003} found a median difference of $-11$\%, opposite to theoretical expectations \citep{Baes2002} and to our results.

It is not immediately clear why \cite{Silge2003} found that NIR velocity dispersion measurements are lower than optical dispersion measurements. One possible reason could be the different spatial range used by \cite{Silge2003}:  Fig.~\ref{dispersionextractionwindow} indicates that velocity dispersions can decrease for higher spatial ranges. The average spatial width in \cite{Silge2003} is $\sim$12$''$, which may be the reason for their difference between NIR and optical velocity dispersions. In order to investigate this possibility, we recalculated the NIR velocity dispersions for all galaxies in our sample, now using an extraction window of $1''\times12''$. The best-fitting line, with a reduced $\chi^2$ of 1.35, is 
\begin{equation}
  \sigma_{\text{opt}}
  = 
  (1.03 \pm 0.07) \sigma_{\text{NIR},1\times12\text{arcsec}^2} 
  -
  (3.35 \pm 9.16),  
\label{eq:rangeSilge} 
\end{equation}
fully consistent with Eq. \ref{eq:corrCOopt}. Fig. \ref{dispersionextractionwindow} suggests an increase of the slope of $\sim 4\%$, which is indeed what we find. Nonetheless, a $\sim 20\%$ effect is needed to explain the results found by \cite{Silge2003}, so we exclude that the extraction width is responsible for the discrepancy. 

Another possible explanation for the discrepancy could be the choice of the templates: Fig.~\ref{EW_sigma_IC1963} showed that the measured velocity dispersion depends strongly on the equivalent width of the CO band of the template. \cite{Silge2003} used templates with equivalent widths ranging from less than 5 to over 20 \AA. So far, we have only used the average K giant template with an equivalent width of 8.99 \AA. In order to investigate this possibility, we have provided the 23 templates we have used in Fig.~\ref{EW_sigma_IC1963} as input templates for pPXF. The resulting velocity dispersions are given in Table \ref{galaxies}. For every galaxy separately, pPXF gives weights to the different templates to obtain the best fit, delivering stellar population information. pPXF favoured HD113538 (KV), HD2490 (MIII) and HD63425B (KIII), which indicates that the light is dominated by cool giants and dwarfs. The best linear fit is given by
\begin{equation}
  \sigma_{\text{opt}}
  = 
  (0.99  \pm   0.07) \sigma_{\text{NIR,23 templates}} 
  + 
  (2.92    \pm   9.05) 
  \label{eq:corr23templ}
\end{equation}
and has a reduced $\chi^2$ of 1.70. It is fully consistent with equation \ref{eq:corrCOopt}, indicating that template mismatch does not affect our velocity dispersion determination. 

\cite{Silge2003} used a nonparametric LOSVD. To investigate whether this could account for the discrepancy, we included the $h_3$ and $h_4$ Gauss-Hermite coefficients, using the average K giant template as input spectrum and the default pPXF parameter $\lambda$ to penalize both parameters. The resulting values for $h_3$ (and $h_4$) are between $-0.05$ and 0.11 ($-0.01$ and 0.08, respectively) with a median value of 0.012 (0.020, respectively). The best-fitting line is given by
\begin{equation}
  \sigma_{\text{opt}}
  = 
  ( 0.99   \pm  0.05) \sigma_{\text{NIR,G-H}} 
  - 
  (2.40   \pm   7.58) \label{eq:h4} 
\end{equation}
with a reduced $\chi^2$ of 0.93. Again, the slope is consistent with 1 and the intercept is consistent with 0, so we exclude that the parametrization of the LOSVD is accountable for the discrepancy.  

One remaining possibility for the difference between our results and \cite{Silge2003} is that the sample selection is important.  Our Fornax sample represents only a cluster environment, while the \cite{Silge2003} sample ranges from cluster to isolated field galaxies and there are known correlations between environment and velocity dispersion (e.g.\ \citealt{Zhu2010,LaBarbera2010}).

\subsection{Correlation with IRAS dust masses}

According to \cite{Silva2008}, at least two galaxies (NGC\,1316 and NGC\,1380) from our sample show clear central dust features. We need to know how much dust there is in the galaxies to study the effects of dust on the observed kinematics. In this section, we derive the dust masses ($M_d$) for our galaxies based on IRAS flux densities at 60 and 100 $\mu$m. We apply the technique of \cite{Goudfrooij1995}, using a distance $D=19.3$~Mpc for all galaxies \citep{Jordan2007}. For 5 galaxies of our sample, both $S_{60}$ and $S_{100}$ were given. In that case both $T_d$ and $M_d$ were determined. If only $S_{100}$ was given, which was the case for 5 galaxies of our sample, we used the average dust temperature $\overline{T}_d=25.9K$ of our galaxy sample as a representative value to compute the dust mass. Table \ref{tabMd} presents the resulting dust masses $M_d$, $T_d$ and the IRAS flux densities for 10 galaxies in our sample. For six galaxies (NGC\,1374, NGC\,1375, NGC\,1381, NGC\,1419, NGC\,1427, ESO\,358-G06) the IRAS Faint Source Catalog lists only upper limits, the three remaining galaxies (NGC\,1373, NGC\,1380A, IC\,1963) are not listed in the catalog.

\begin{table*}
\caption{\label{tabMd} IRAS flux densities and dust characteristics.}
\begin{tabular}{lccccc}
\hline
Galaxy & S($60 \mu$m)  & S($100 \mu$m) & log($M_d$) & $T_d$ & log($M_d/L_B$)\\
&[Jy]& [Jy] & [$M_\odot$] & [K]&[$M_\odot/L_\odot$]\\
\hline
NGC\,1316 & 3.070 $\pm$ 0.030 &8.110 $\pm$ 1.900 &6.11 $\pm$ 0.25&27.2 $\pm$ 2.2&$-4.64$ \\
NGC\,1336 &... &0.260 $\pm $ 0.095&4.73&...&$-4.54$\\
NGC\,1339&0.2298 $\pm$ 0.14&0.670  $\pm$ 0.058 &5.09 $\pm$ 0.25&26.5  $\pm$ 2.1&$-4.42$\\
NGC\,1351&0.090  $\pm$0.021&0.510  $\pm$ 0.042&5.41 $\pm$ 0.24&22.4  $\pm$ 1.5&$-4.10$\\
NGC\,1379&...&0.140  $\pm$ 0.047 &4.46&...&$-5.33$\\
NGC\,1380&1.040 $\pm$ 0.042&3.440 $\pm$ 0.107&5.88 $\pm$ 0.24&25.6  $\pm$ 1.9&$-4.27$\\
NGC\,1399&...&0.300 $\pm$ 0.082&4.79&...&$-5.48$\\
NGC\,1404&...&0.270 $\pm$ 0.056&4.75&...&$-5.36$\\
IC\,2006&0.120 $\pm$0.015&0.320 $\pm$ 0.047&4.71 $\pm$0.25&27.2 $\pm$2.2&$-4.92$\\
ESO\,358-G59&...&0.380 $\pm$ 0.060&4.90&...&$-4.01$\\
\hline
\end{tabular}
\end{table*}

\begin{figure}
\centering
\includegraphics[width=0.45\textwidth]{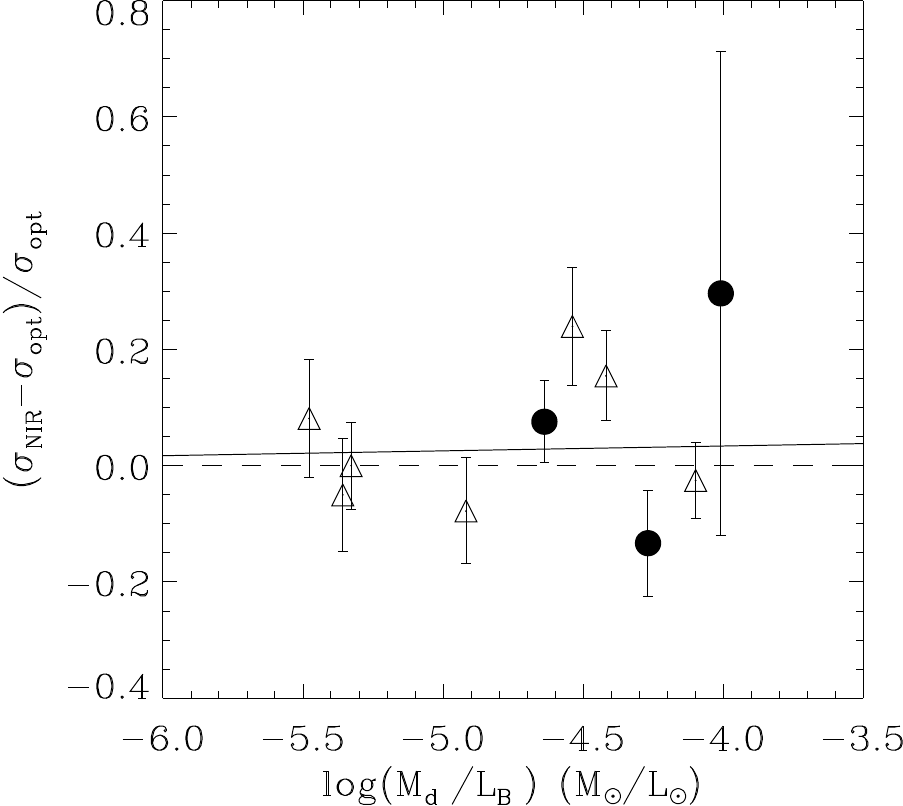}
\caption{Fractional difference between the infrared and optical dispersions as a function of the ratio of dust mass to $B$-band luminosity. The solid line is the best fit, given by Eq. \ref{eq:fracdiffMd}. The Spearman rank-order correlation coefficient is 0.21 at a significance level of 0.56. } 
\label{figfracdiffMd}
\end{figure}

Fig.~\ref{figfracdiffMd} presents the fractional difference of the infrared and the optical dispersions as a function of the relative amount of dust in a galaxy, which is estimated in \cite{Silge2003} by the ratio of the IRAS dust mass to the $B$-band luminosity. The estimates for the relative amount of dust are given in Table \ref{tabMd}. Note that the IRAS dust mass estimates are a lower limit for the true dust masses, because IRAS is not sensitive to cold dust (which emits the bulk of its radiation longwards of 100$\mu$m.). A negligible trend is visible (confirmed by the Spearman rank-order correlation coefficient equal to 0.21).
The best-fitting solid line is given by equation
\begin{equation}
  \frac{\sigma_{\text{NIR}}-\sigma_{\text{opt}}}{\sigma_{\text{opt}}}
  = 
  (0.008 \pm 0.056) \log \left(\frac{M_d}{L_B}\right)+ (0.068 \pm  0.266),
\label{eq:fracdiffMd}
\end{equation}
with both the slope and the intercept consistent with 0 and a reduced $\chi^2$ of 1.76. This implies that warm dust does not affect optical dispersions. We cannot yet make the same conclusion for colder dust, but Fig. \ref{corrCOoptical} indicates that the typical effect is very weak. 

\section{Conclusions} \label{sec:conclusions} 

In this study, we investigate a complete magnitude-limited and unbiased sample of 22 early-type galaxies in the Fornax cluster and are able to determine the kinematics based on the 2.29$\mu$m $^{12}$CO(2-0) feature for 19 of those galaxies. We related the NIR velocity dispersions with the optical dispersions of \cite{Kuntschner2000} and found no evidence for a $\sigma$-discrepancy for the ellipticals nor for the lenticulars. Our results agree with a previous smaller study of \cite{Silva2008}, but not with \cite{Silge2003}. We investigated this disagreement by providing a variety of input templates to pPXF with a large range of EWs, by changing the spatial width of the extraction window and by introducing Gauss-Hermite coefficients in the LOSVD, but we were not able to clarify this discrepancy.

We have computed the dust masses based on IRAS flux densities for 10 galaxies of our sample and investigated the influence of diffuse dust on the observed kinematics, which turned out to be negligible. 

The one-to-one correspondence between the optical and the NIR velocity dispersions found for this homogeneous set of early-type galaxies implies that velocity dispersions measured at optical wavelengths are reliable kinematic parameters for early-type galaxies and hence that no bias is introduced in statistical relations that build on such dispersions (such as the $\text{M}_{\text{BH}}$-$\sigma$ relation or the Fundamental Plane). Combined with the simulations by \cite{Baes2000,Baes2002}, it also supports the traditional point of view on the dust content of early-type galaxies, namely that they are virtually optically thin. While some observational studies hinted towards the existence of a substantial diffuse dust component in early-type galaxies (e.g. \citealt{Temi2004,Temi2007,Leeuw2004,Vlahakis2005}), the most recent results from the recently launched Herschel Space Observatory indicate a dearth of diffuse dust in the few elliptical galaxies studied so far (\citealt{Clemens2010,Baes2010,Gomez2010}). 

Whereas our results support this scenario, we must be careful for two caveats. On the one hand, our results only set limits on the presence of a smooth, diffusely distributed dust component, which one would expect if the dust has an internal origin. If the dust has an external origin, it is not necessarily coincident with the stellar body. An example is the nearby Virgo Cluster elliptical M86, where the dust is clearly related to stripping from a nearby spiral galaxy and is concentrated some 10 kpc to the south-east of the nucleus \citep{Gomez2010}.

On the other hand, the comparison of optical and NIR velocity dispersions might not be the most sensitive way to measure the optical thickness of early-type galaxies. The simulations of \cite{Baes2000,Baes2002} indicate an effect of a few percent only for optical depths of order unity. Combined with the measurement errors and other possible effects such as different stellar populations dominating the kinematics at optical and NIR wavelengths, our results should be considered only in a statistical sense and one should take care not to directly interpret results on individual galaxies in terms of optical thickness. A clear example is the case of NGC\,1380: in spite of a clear dust lane in optical images \citep{Jordan2007} and a significant IRAS dust mass of $7.6 \times 10^5~M_\odot$, the NIR velocity dispersion is lower than the optical dispersion.

A clearer picture on the optical thickness of early-type galaxies will hopefully emerge in the near future when substantial numbers of nearby early-type galaxies will be observed as part of several Herschel Key Programs, including the Herschel Virgo Cluster Survey \citep{Davies2010} and the Herschel Reference Survey \citep{Boselli2010}.

\section*{Acknowledgments}

This research has made use of NASA's Astrophysics Data System and the NASA/IPAC Extragalactic Database (NED) which is operated by the Jet Propulsion Laboratory, California Institute of Technology, under contract with the National Aeronautics and Space Administration. This research also has made use of the SIMBAD database, operated at CDS, Strasbourg, France.  JV acknowledges FWO Vlaanderen for the financial support.  AJR was supported by the FONDAP Center for Astrophysics CONICYT 15010003 and by the National Science Foundation grants AST-0808099 and AST-0909237. We also wish to thank the referee for constructive comments on the manuscript.

\bibliographystyle{mnras}
\bibliography{references}
\label{lastpage}

\end{document}